\documentclass[pra,twocolumn,amsmath,amssymb,showpacs]{revtex4}
\newcommand{\ocis}[1]{}
\newcommand{\citeonline}[1]{\cite{#1}}

\usepackage[draft]{hyperref} 

\usepackage{amsmath}
\usepackage{amssymb}
\usepackage{bbm}

\newcommand{\be}{\begin{equation}}
\newcommand{\ee}{\end{equation}}
\newcommand{\eea}{\end{eqnarray}}
\newcommand{\bea}{\begin{eqnarray}}

\newcommand{\va}[1]{\ensuremath{(\Delta#1)^2}}

\newcommand{\ex}[1]{\ensuremath{\left\langle{#1}\right\rangle}}
\newcommand{\exs}[1]{\ensuremath{\langle{#1}\rangle}}

\newcommand{\qed}{\ensuremath{\hfill \Box}}

\newcommand{\ketbra}[1]{\ensuremath{| #1 \rangle \langle #1 |}}

\newcommand{\ket}[1]{\ensuremath{|#1\rangle}}

\newcommand{\kommentar}[1]{}

\begin{document}

\title{Detection of multipartite entanglement in the vicinity of symmetric Dicke states}

\author{G\'eza T\'oth}

\email{toth@alumni.nd.edu}

\affiliation{Max Planck Institute for Quantum
Optics, Hans-Kopfermann-Stra{\ss}e 1, D-85748 Garching, Germany,}
\affiliation{Research Institute for Solid State Physics and Optics,
Hungarian Academy of Sciences,  P.O. Box 49, H-1525 Budapest,
Hungary}

\begin{abstract}We present methods for detecting entanglement around symmetric Dicke states.
In particular, we consider $N$-qubit symmetric Dicke states with
$N/2$ excitations. In the first part of the paper we show that for
large $N$ these states have the smallest overlap possible with
states without genuine multi-partite entanglement. Thus these states
are particulary well suited for the experimental examination of
multi-partite entanglement. We present fidelity-based entanglement
witness operators for detecting multipartite entanglement around
these states. In the second part of the paper we consider
entanglement criteria, somewhat similar to the spin squeezing
criterion, based on the moments or variances of the collective spin
operators. Surprisingly, these criteria are based on an upper bound
for variances for separable states. We present both criteria
detecting entanglement in general and criteria detecting only
genuine multi-partite entanglement. The collective operator measured
for our criteria is an important physical quantity: Its expectation
value essentially gives the intensity of the radiation when a
coherent atomic cloud emits light.
\end{abstract}


\ocis{270.0270, 270.4180, 999.9999 multipartite quantum entanglement}

\maketitle 

\section{Introduction}

The nonclassical effects of quantum mechanics have already been
studied theoretically for more than 50 years years \cite{EPR}. Which
quantum states can lead to phenomena that are strikingly
nonclassical? Which quantum states are useful for quantum
information processing applications? The answers to these questions
lead to the definition of separability, entanglement \cite{W89}, and
multi-partite entanglement \cite{SK00,AB01}.

In the last decade, with the rapid development of quantum control
\cite{BE00} it has become possible to examine the nonclassicality of
quantum mechanics experimentally by creating multi-qubit quantum
states of photons
\cite{ZC04,PB00,PB00B,ZY03,BE04,clusterexp,clusterexp2,clusterexp3},
trapped ions \cite{SK00}, and cold atoms on an optical lattice
\cite{MG03B}. The first multi-qubit experiments concentrated on
Greenberger-Horne-Zeilinger \cite{GH90} (GHZ)  states. As maximally
entangled multi-qubit states, they are intensively studied and have
been realized in numerous experiments \cite{PB00,PB00B,ZY03,SK00}.
Other quantum states targeted in experiments due to their
interesting properties are, for example, cluster states
\cite{BR03,RB03,clusterexp,clusterexp2,clusterexp3} and many-body
singlet states \cite{singlet}.

In this paper we will discuss some of the advantages of using Dicke
states \cite{Dicke} to study the nonclassical phenomena of quantum
mechanics. In his seminal paper Ref.~\citeonline{Dicke}, Dicke
considered the spontaneous emission of light by a cloud of two-state
atoms which are coupled to the electromagnetic field as electric
dipoles. He found that when the cloud acts as a coherent quantum
system, the maximal light intensity is roughly proportional to the
square of the number of atoms. This Dicke called superradiance. The
highly correlated Dicke states, defined to describe the system
above, are the simultaneous eigenstates of the collective angular
momentum, $J$ and its $z$-component, $J_z$. In a typical many-qubit
experiment, in which the qubits cannot be individually accessed,
both the initial state and the dynamics are symmetric under the
permutation of qubits. Thus in this paper we will consider only
symmetric Dicke states. These are also the states with maximal $J.$
An $N$-qubit symmetric Dicke state with $m$ excitations is defined
as \cite{SG03}
\begin{equation}
\ket{m,N}:=\bigg(\begin{array}{c}N \\
m\end{array}\bigg)^{-\frac{1}{2}}\sum_k P_k
(\ket{1_1,1_2,...,1_m,0_{m+1},...,0_N}), \label{sd}
\end{equation}
where $\{P_k\}$ is the set of all distinct permutations of the
spins. $\ket{1,N}$ is the well known $N$-qubit W state.

Several proposals have been presented in the literature for the
experimental creation of Dicke states. In Ref.~\citeonline{UF03} a
scheme is considered for creating Dicke states in trapped ions using
an adiabatic process. A method for the realization of arbitrary
superposition of symmetric Dicke states by detecting the photons
leaving a cavity is described in Ref.~\citeonline{DK03}. A novel
scheme has been proposed for obtaining Dicke states based on
creating closed subspaces for the quantum dynamics of an ion chain
\cite{Enrique}. Other proposals are described for example in
Refs.~\citeonline{SH04,ZP02,XH04,GT05}.

On the experimental side, we have to mention that a three-qubit W
state has been created in a photonic system \cite{EK04,BE04,ML05}.
Also, an eight-qubit W state has been prepared with trapped ions
\cite{OGNature}. Very recently, a four-qubit Dicke state with two
excitations has been created in a photonic system \cite{Experiment}.
It turned out that this is one of the quantum states which can be
obtained in a photonic experiment with a very good fidelity. Future
experiments will most certainly lead to creation of Dicke states
with multiple excitations in other physical systems. At this point
it is important to ask the question: Are such states more useful
than others from the point of view of quantum information
processing? In Ref.~\citeonline{Experiment} it has already been
discussed that the Dicke state prepared in the experiment is useful
for telecloning.

In this paper we demonstrate that Dicke states with multiple
excitations are also good candidates for the experimental
examination of genuine multipartite entanglement \cite{AB01}. In
particular, we discuss how to detect entanglement close to
$\ket{N/2,N}$, i.e., an $N$-qubit symmetric Dicke state with $N/2$
excitations. We find that, similarly to GHZ \cite{GH90} and cluster
states \cite{BR03}, for large $N$ such states have the smallest
overlap possible with states without multipartite entanglement.

In the second part of the paper entanglement detection schemes
requiring only collective measurements are discussed
\cite{SM01,SM01_2,SM01_3,SM01_4,T04,KC05,KC05_2}. Entanglement
detection with collective measurements is important since in many
experiments the qubits cannot be accessed individually. Even if the
qubits can be individually accessed, our measurement schemes are
still useful since they need a small experimental effort
\cite{Experiment}. The schemes presented are based on an {\it upper}
bound on the variances of collective observables for separable
states. Any state violating this bound is detected as entangled. We
present schemes for entanglement detection in general and also
schemes for detecting only genuine multi-partite entanglement.

$\ket{N/2,N}$ is exactly the quantum state for which Dicke found
that the superradiance is the strongest \cite{Dicke} for even $N.$
We will show that if our schemes are applied to a system described
in Dicke's original paper \cite{Dicke} then the measurement of the
collective observables of our scheme is essentially equivalent to
the measurement of light intensity emitted by the atoms.

Our paper is organized as follows. In Sec.~II we show that for a
fidelity-based detection of multipartite entanglement the required
fidelity is low for this state. In Sec.~III. we discuss entanglement
detection with collective observables close to the state
$\ket{N/2,N}.$ In the Appendix we present some calculations for
Sec.~II.

\section{Fidelity-based entanglement criteria}

Before starting our main discussion, let us first review the basic
terminology of the field. An $N$-qubit state is called fully
separable if its density matrix can be written as the mixture of
product states
\begin{equation}
\rho = \sum_l p_l \rho_l^{(1)} \otimes \rho_l^{(2)} \otimes
...\otimes \rho_l^{(N)}, \label{sep}
\end{equation}
where $\sum_l p_l=1$ and $p_l>0.$ Otherwise the state is called
entangled. Quantum optics experiments aim to create entangled
states, since these are the quantum states which lead to phenomena
very different from classical physics \cite{BE00}.

In a multi-qubit experiment it is important to detect genuine
multi-qubit entanglement \cite{BE04}: We have to show that all the
qubits were entangled with each other, not only some of them. An
example of the latter case is a state of the form
\begin{equation}
\ket{\Phi}=\ket{\Phi_{1..m}}\otimes\ket{\Phi_{m+1..N}} \label{Psi}
\end{equation}
Here $\ket{\Phi_{1..m}}$ denotes the state of the first $m$ qubits
while $\ket{\Phi_{m+1..N}}$ describes the state of the remaining
qubits. Note that the state given by Eq.~(\ref{Psi}) might be
entangled, but it is separable with respect to the partition
$(1,2,..,m)(m+1,m+2,..,N).$ Such states are called biseparable
\cite{AB01} and can be created from product states such that two
groups of qubits do not interact. These concepts can be extended to
mixed states. A mixed state is biseparable if it can be created by
mixing biseparable pure states of the form Eq.~(\ref{Psi}). Note
that we get mixed biseparable states even when mixing pure
biseparable states which are separable with respect to different
partitions (e.g., when mixing $(\ket{00}+\ket{11})\ket{0}/\sqrt{2}$
and $\ket{0}(\ket{00}+\ket{11})/\sqrt{2}$). An $N$-qubit state is
said to have genuine $N$-partite entanglement if it is not
biseparable.

Now we will present conditions for the detection of genuine
multipartite entanglement. These will be criteria based on
entanglement witness operators
\cite{entwit,entwit1,entwit2,entwit3,entwit4,entwit5,entwit6,entwit7}.
In other words, these are criteria which are linear in operator
expectation values \cite{nonlin}. Based on Ref.~\citeonline{BE04} we
know that for biseparable states $\rho$
\begin{equation}
Tr(\rho\ketbra{\Psi})\le C_\Psi. \label{cond}
\end{equation}
Here $\ket{\Psi}$ is a multi-qubit entangled state and $C_\Psi$ is
the square of the maximal overlap of $\ket{\Psi}$ with biseparable
states \cite{BE04}
\begin{equation}
C_\Psi:=\max_{\phi\in \mathcal{B}} |\exs{\Psi|\phi}|^2, \label{C}
\end{equation}
where $\mathcal{B}$ denotes the set of biseparable pure states. Any
state $\rho$ violating Eq.~(\ref{cond}) is necessarily genuine
multipartite entangled. The bound in Eq.~(\ref{cond}) is sharp, that
is, it is the lowest possible bound. Computing Eq.~(\ref{C}) seems
to be a complicated optimization problem. Fortunately, it turns out
that $C_\Psi$ equals the square of the maximum of the Schmidt
coefficients of $\ket{\Psi}$ with respect to all bipartitions
\cite{BE04}. Thus $C_\Psi$ can be determined easily, without the
need for multi-variable optimization.

The use of criteria of the type Eq.~(\ref{cond}) are the following.
Let us say that in an experiment one aims to prepare the state
$\ket{\Psi}.$ This preparation is not perfect, however, one might
still expect that the state prepared in the experiment is close to
$\ket{\Psi}.$ Thus a fidelity-based entanglement criterion of the
type Eq.~(\ref{cond}) can be used to detect its entanglement. The
smaller the required minimal fidelity $C_\Psi$, the better the
criterion from practical point of view.

Now we will present criteria of the form Eq.~(\ref{cond})
for detecting entanglement around symmetric Dicke states.\\
{\bf Theorem 1.} For biseparable quantum states $\rho$
\begin{equation}
Tr(\rho\ketbra{N/2,N})\le \frac{1}{2}\frac{N}{N-1}=:C_{N/2,N}.
\end{equation}
This condition detects entanglement close to
an $N$-qubit symmetric Dicke state with $N/2$ excitations. Here $N$ is assumed to be even. \\
{\it Proof.} The Schmidt decomposition of $\ket{m,N}$ according to
the partition $(1,2, ... ,N_1) (N_1+1,N_1+2, ... ,N)$ is \cite{SG03}
\begin{equation}
\ket{m,N}= \sum_k \lambda_k \ket{k,N_1} \otimes \ket{m-k,N-N_1},
\end{equation}
where the Schmidt coefficients are
\begin{equation}
\lambda_k=\bigg(\begin{array}{c}N \\
m\end{array}\bigg)^{-\frac{1}{2}} \bigg(\begin{array}{c}N_1\\
k\end{array}\bigg)^{\frac{1}{2}} \bigg(\begin{array}{c}N-N_1 \\
m-k\end{array}\bigg)^{\frac{1}{2}}.
\end{equation}
We do not have to consider other partitions due to the permutational
symmetry of our Dicke states. For $\ket{N/2,N}$ we have $m=N/2.$ Now
we use that
\begin{equation}
\bigg(\begin{array}{c}N_1 \\ k\end{array}\bigg)
\bigg(\begin{array}{c}N-N_1 \\ \frac{N}{2}-k\end{array}\bigg) \le
\bigg(\begin{array}{c} 2 \\ 1\end{array}\bigg)
\bigg(\begin{array}{c} N-2 \\ \frac{N}{2}-1\end{array}\bigg).
\label{ineq}
\end{equation}
The proof of Eq.~(\ref{ineq}) can be found in the Appendix. Thus we
find that the maximal Schmidt coefficient can be obtained for
$N_1=2$ and $k=1.$ For $N_1=2$ we obtain $\lambda_1^2=N(N-1)/2.$
$\qed$

Thus we find that $C_{N/2,N}\approx 1/2$ for large $N.$ This makes
the detection of multipartite entanglement around the state
$\ket{N/2,N}$ relatively easy. This property is quite remarkable: Up
to now only GHZ \cite{GH90}, cluster \cite{BR03} and graph states
\cite{HE04} known to have $C=1/2$ \cite{TG05,TG05_2}.

Connected to the previous paragraph, it is important to check how
much our entanglement criterion is robust against noise. In order to
see that let us consider a $\ket{N/2,N}$ state mixed with white
noise:
\begin{equation}
\varrho(p)=p_{\rm noise} \frac{\openone}{2^N} + (1-p_{\rm noise})
\ketbra{N/2,N}, \label{noise}
\end{equation}
where $p_{\rm noise}$ is the ratio of noise. Our criterion is very
robust: It detects a state of the form Eq.~(\ref{noise}) as true
multipartite entangled if
\begin{equation}
p_{\rm noise} < \frac{1}{2}\bigg[\frac{N-2}{(N-1)(1-2^{-N})}\bigg].
\end{equation}
For large $N$ we have $p_{\rm noise} \le 1/2.$

Note that the situation is very different for a W state. A condition
which can be obtained for detecting genuine multi-partite
entanglement around a W state is \cite{BE04,OGNature}
\begin{equation}
Tr(\rho\ketbra{1,N})\le \frac{N-1}{N}=:C_{1,N}. \label{Wcrit}
\end{equation}
Any state violating this condition is multi-partite entangled.
However, note that with an increasing $N$, $C_{1,N}$ approaches
rapidly $1.$ This makes multipartite entanglement detection based on
Eq.~(\ref{Wcrit}) challenging.

\section{Entanglement detection with collective measurements}

In Sec. II. for Theorem 1 we needed the measurement of the
expectation value of $\ketbra{N/2,N}.$ In order to measure this
operator, it must be decomposed into the sum of multi-qubit
correlation terms of the form $A^{(1)}\otimes A^{(2)} \otimes
A^{(3)} \otimes ... $
\cite{decomposition,decomposition2,decomposition3,decomposition4,BE04,TG05,TG05_2},
where $A^{(k)}$ acts on qubit $k.$ For measuring the expectation
value of such correlation terms, we must be able to access the
qubits individually.

However, in certain physical systems (e.g., optical lattices of
bosonic two-state atoms \cite{MG03B}) only the measurement of
collective quantities is possible. In this section we present
entanglement criteria for detecting entanglement with collective
measurements \cite{SM01,SM01_2,SM01_3,SM01_4,T04,KC05,KC05_2}. Our
entanglement conditions will be built using the collective spin
operators
\begin{equation}
J_{x/y/z}:=\frac{1}{2}\sum_{k=1}^N \sigma_{x/y/z}^{(k)},
\end{equation}
where $\sigma_{x/y/z}^{(k)}$ denote Pauli spin matrices acting on
qubit $k.$

 {\bf Lemma 1.} For separable states the maximum of the
expression
\begin{equation}
a_x\exs{J_x^2}+a_y\exs{J_y^2}+a_z\exs{J_z^2}+
b_x\exs{J_x}+b_y\exs{J_y}+b_z\exs{J_z} \label{min}
\end{equation}
with $a_{x/y/z}\ge 0$ and real $b_{x/y/z}$ is the same as its
maximum for translationally invariant product states (i.e., for
product states of the form $\ket{\Psi}=\ket{\psi}^{\otimes N}$). In
particular, if $b_x=b_y=b_z=0$ then this expression is bounded from
above by
\begin{equation}
B:=(a_x+a_y+a_z)\frac{N}{4}+\max(a_x,a_y,a_z)\frac{N}{2}
\left(\frac{N}{2}-\frac{1}{2}\right).\label{Bf}
\end{equation}
\\
{\it Proof.} Due to the convexity of the set of separable states, it
is enough to look for the maximum for pure product states. For
technical reasons, let us consider a mixed product state of the form
$\rho=\otimes_{k=1}^N \rho^{(k)} $ and use the notation
$s_{x/y/z}^{(k)}:=Tr(\rho^{(k)}\sigma_{x/y/z} )/2.$ Hence we have to
maximize
\begin{eqnarray}
f:&=&(a_x+a_y+a_z)N\nonumber\\&+&\sum_{l=x,y,z} a_l \left[\left(\sum_{k}
s_{l}^{(k)}\right)^2-\sum_k \left(s_{l}^{(k)}\right)^2\right]\nonumber\\
&+&b_l \sum_{k} s_{l}^{(k)}.\label{feq}
\end{eqnarray}
Let us consider the constraints
\begin{equation}
\sum_{k} s_{l}^{(k)}=K_l \label{cons}
\end{equation}
for $l=x,y,z$ where $K_l$ are some constants. Note that $f$ can be
written as $f=(a_x+a_y+a_z)N+a_xf_x+a_yf_y+a_zf_z.$ Now let us first
take $f_x$, that is, the part which depends only on the
$s_{x}^{(k)}$ coordinates. It can be written as
\begin{equation}
f_x=\left(\sum_{k}
s_{x}^{(k)}\right)^2-\sum_k \left(s_{x}^{(k)}\right)^2+\alpha\sum_{k}
s_{x}^{(k)},
\label{fx}
\end{equation}
where $\alpha_x=b_x/a_x.$ We build the constraint Eq.~(\ref{cons})
into our calculation by the substitution
\begin{equation}
s_{x}^{(N)}=K_x-\sum_{k=1}^{N-1} s_{x}^{(k)}.
\end{equation}
Then for any $m<N$ we obtain the derivatives as
\begin{equation}
\frac{\partial f_x}{\partial
s_{x}^{(m)}}=-2s_{x}^{(m)}+2(K_x-\sum_{k=1}^{N-1} s_{x}^{(k)}).
\end{equation}
In an extreme point this should be zero. Hence it follows that for
all $m<N$
\begin{equation}
s_{x}^{(m)}=s_{x}^{(N)},
\end{equation}
thus $f_x$ takes its extremum when all $s_x^{(m)}$ are equal. Let us
now see whether this extreme point is a maximum. For any $m,n<N$
\begin{equation}
\frac{\partial^2 f_x}{\partial s_{x}^{(m)} \partial
s_{x}^{(n)}}=-2-2\delta_{mn},
\end{equation}
where $\delta_{mn}$ is the Kronecker symbol. It is easy to see, that
the matrix containing the second order derivatives is negative
definite, thus our extremum is a maximum. It is also a global
maximum, since based on Eq.~(\ref{fx}) and the constraint
Eq.~(\ref{cons}) it is obvious that if any $|s_x^m|\rightarrow
\infty$ then $f_x\rightarrow - \infty.$ Similar calculations can be
carried out for the part of $f$ depending on the $y$ and $z$
coordinates. We have just proved that for given $K_{x/y/z}$, $f$
given in Eq.~(\ref{feq}) takes its maximum for translationally
invariant product states for which $s_{x/y/z}^{(k)}=K_{x/y/z}/N.$
This maximum we will denote by $f_{\rm max}(K_x,K_y,K_z).$

Let us now look for the $K_x$, $K_y$ and $K_z$ for which $f_{\rm
max}$ maximal. The condition for getting a physical state is $\sum_l
(K_{l}/N)^2\le 1/4$ where the equality holds for pure product
states. We find that $f_{\rm max}$ is a convex function thus it
takes its maximum at the boundary of the domain allowed for
$K_{x/y/z},$ i.e., it takes its maximum for pure translationally
invariant product states. Hence the upper bound Eq.~(\ref{Bf}) for
$f$ follows. $\qed$

In general it is very hard to find the maximum for an operator
expectation value for separable states
\cite{EH04,EH04_2,EH04_3,EH04_4}. We have just proved that for
operators of the form Eq.~(\ref{min}) which are constructed from
first and second moments of the angular momentum coordinates this
problem is easy: It can be reduced to a maximization over states of
the form $\ket{\psi}^{\otimes N},$ i.e., to a maximization over
three real variables $s_{x/y/z}.$ Note that it is not at all clear
from the beginning that this simplification is possible. For
example, when looking for the minimum of $J_x^2+J_y^2+J_z^2$ for
pure product states, it turns out that the expression is minimized
not by translationally invariant product states. To be more
specific, for $N=2$ when we minimize this expression for product
states, the minimum is obtained for the state $\ket{1}\ket{-1}.$

{\bf Theorem 2.} As a special case of the previous criterion, we
have that for separable states \cite{moregeneral}
\begin{equation}
\exs{J_x^2}+\exs{J_y^2}\le
\frac{N}{2}\left(\frac{N}{2}+\frac{1}{2}\right). \label{crit}
\end{equation}
For even $N$, the left hand side is the maximal
$\frac{N}{2}\left(\frac{N}{2}+1\right)$ only for an $N$-qubit
symmetric Dicke state with $N/2$ excitations. Based on Lemma 1, the
proof of this theorem is obvious. It can also be seen that the bound
in Eq.~(\ref{crit}) is sharp since a separable state of the form
\begin{equation}
\ket{\Psi_{\rm xy}}:=2^{-N/2}(\ket{0}+\ket{1}e^{i\phi})^{\otimes N}
\label{psixy}
\end{equation}
for any real $\phi$ saturates the bound.

Based on Eq.~(\ref{crit}), it is easy to see that for separable
states we also have
\begin{equation}
\va{J_x}+\va{J_y}\le
\frac{N}{2}\left(\frac{N}{2}+\frac{1}{2}\right). \label{crit_va}
\end{equation}
Thus $J_{x/y}^2$ could be replaced by the corresponding variances.
Any state violating Eq.~(\ref{crit_va}) is entangled. Note the
curious nature of our criterion: A state is detected as entangled,
if the uncertainties of the collective spin operators are {\it
larger} than a bound.

How can we intuitively understand the criterion Eq.~(\ref{crit})?
Using the notation $\vec{J}=(J_x,J_y,J_z)$, one can rewrite it as
\cite{SolPrivate}
\begin{equation}
\exs{\vec{J}^2}- \frac{N}{2}\left(\frac{N}{2}+\frac{1}{2}\right) \le
\exs{J_z^2}. \label{critJ}
\end{equation}
For a given $\exs{\vec{J}^2}$, in order to violate
Eq.~(\ref{critJ}), $\exs{J_z^2}$ must be sufficiently low. For
symmetric states (i.e., for states which could be used for
describing two-state bosons) we have
$\exs{\vec{J}^2}=\frac{N}{2}\left(\frac{N}{2}+1\right)$ and
Eq.~(\ref{critJ}) turns into the condition
\begin{equation}
\frac{N}{4} \le \exs{J_z^2}. \label{critJ2}
\end{equation}
A condition similar to Eq.~(\ref{critJ2}) has already been presented
for the detection of two-qubit entanglement for symmetric states in
Refs.~\citeonline{KC05,KC05_2}.

Criterion Eq.~(\ref{crit}) detects the state of the form
Eq.~(\ref{noise}) as entangled if $p_{\rm noise}<1/N.$ Note that the
limit on $p_{\rm noise}$ decreases rapidly with $N.$ Let us now
consider a different type of noise:
\begin{eqnarray}
\varrho'(p)&=&p_{\rm noise} \ketbra{\Psi_{\rm xy}} \nonumber\\
&+&(1-p_{\rm noise}) \ketbra{N/2,N},
\end{eqnarray}
where $\Psi_{\rm xy}$ is defined in Eq.~(\ref{psixy}). Then
criterion Eq.~(\ref{crit}) detects the state as entangled for any
$p_{\rm noise}<1.$ Thus the usefulness of our criteria depends
strongly on the type of the noise appearing in an experiment.

Criteria can also be obtained which detect entanglement around other
multi-qubit Dicke states. For example, the expression \cite{UF03}
\begin{equation}
\exs{J_x^2}+\exs{J_y^2}-2m\exs{J_z}\label{crit2}
\end{equation}
takes its maximum at a Dicke state $\ket{m+N/2,N}.$ The maximum for
separable states can be obtained from Lemma 1.

Up to now we discussed how to detect entanglement with the
measurement of collective observables. Now we show that a criterion
similar to the one in Theorem 2 can be used to detect genuine
multipartite entanglement. Such a criterion has already been
presented for three qubits in Ref.~\citeonline{AppC}. For
biseparable three-qubit states
\begin{equation}
\exs{J_x^2}+\exs{J_y^2} \le 2 + \sqrt{5}/2\approx 3.12.
\label{crit3q}
\end{equation}
Both the state $\ket{W}=\ket{1,3}$ and the state
$\ket{\overline{W}}=\ket{2,3}$ give the maximal $3.75$ for the
left-hand side of Eq.~(\ref{crit3q}).

Now let us look for criteria for larger systems.
In order to proceed, we will need the following:\\
{\bf Lemma 2.} For a two-qubit quantum state
\begin{equation}
\exs{M_1}^2+\exs{M_2}^2+\exs{M_3}^2\le \frac{16}{3}, \label{Lemma}
\end{equation}
where
\begin{eqnarray}
M_1&:=&\sigma_x^{(1)}\sigma_x^{(2)}+\sigma_y^{(1)}\sigma_y^{(2)},\nonumber\\
M_2&:=&\sigma_x^{(1)}+\sigma_x^{(2)},\nonumber\\
M_3&:=&\sigma_y^{(1)}+\sigma_y^{(2)}.
\end{eqnarray}
{\it Proof.} The proof is rather technical. Let us consider the
vector $v:=(\exs{M_1},\exs{M_2},\exs{M_3}).$ We want to find an
upper bound on $|v|.$ We can easily write
\begin{equation}
|v|^2=\exs{M_1}^2+\exs{M_2}^2+\exs{M_3}^2.
\end{equation}
We have to look for the maximum of this expression for quantum
states. The problem is that it is nonlinear in operator expectation
values. Because of that we will employ the following equality
\begin{eqnarray} |v|&=&\max_{|n|=1} vn, \label{v1}
\end{eqnarray}
where $n$ is a real unit vector. The meaning of Eq.~(\ref{v1}) is
clear: The length of a vector equals to the maximum of its scalar
product with a unit vector. Now the right hand side of
Eq.~(\ref{v1}) can be rewritten as
\begin{eqnarray}
|v| = \max_{|n|=1} \exs{n_1M_1+n_2M_2+n_3M_3}. \label{v2}
\end{eqnarray}
The advantage of this expression is that it is linear in operator
expectation values. The disadvantage is that we have to maximize
over $n.$ Now we will find an upper bound on the right hand side of
Eq.~(\ref{v2}). We will use the fact that for an operator $A$ the
expectation value is bounded as $\exs{A}\le \Lambda_{\max}(A).$ Here
$\Lambda_{\max}(A)$ denotes the largest eigenvalue of operator $A.$
Thus
\begin{eqnarray}
|v|\le \max_{|n|=1} \Lambda_{\max} (n_1M_1+n_2M_2+n_3M_3). \label{v}
\end{eqnarray}
The eigenvalues of $(n_1M_1+n_2M_2+n_3M_3)$ can easily be obtained
analytically as the function of $n_k.$ They are
\begin{eqnarray}
\lambda_1&=&0,\nonumber\\
\lambda_2&=&-2n_1,\nonumber\\
\lambda_{3/4}&=&n_1\pm\sqrt{n_1^2+4n_2^2+4n_3^2}. \label{eig2}
\end{eqnarray}
Assuming $|n|=1$, the eigenvalues given in Eq.~(\ref{eig2}) are
bounded from above by $\sqrt{16/3}.$ Hence, based on Eq.~(\ref{v})
we obtain $|v|^2\le 16/3$ and Eq.~(\ref{Lemma}) follows. $\qed$

Using Lemma 2, we can state the following:\\
{\bf Theorem 3.} For a four-qubit biseparable state
\begin{equation}
\ex{J_x^2}+\ex{J_y^2} \le \frac{7}{2}+\sqrt{3}\approx 5.23
\label{4qubit}
\end{equation}
For the left hand side of Eq.~(\ref{4qubit}) the maximum is $6$ and
it is obtained uniquely for the $\ket{2,4}$ state.\\ {\it Proof.}
First we present the proof for biseparable pure states with a
$(12)(34)$ partition. For these $\ex{J_x^2}+\ex{J_y^2}=2+v_1v_2/2$
where
\begin{align}
v_1&:=( & x_1x_2+y_1y_2, \; & x_1+x_2, \; & y_1+y_2, \; & 1
&),\nonumber\\
v_2&:=(& 1,  \; & x_3+x_4, \; & y_3+y_4, \; & x_3x_4+y_3y_4 &).
\end{align}
Here we used the notation
$x_1x_2=\exs{\sigma_x^{(1)}\sigma_x^{(2)}}$ Hence a bound can be
obtained using the Cauchy-Schwarz inequality as
$\ex{J_x^2}+\ex{J_y^2}\le 2+|v_1||v_2|/2\le 31/6\approx 5.17,$ where
we used that due to Lemma 2 we have $|v_k|^2\le 16/3.$ Note that the
upper bound we have just obtained for $\ex{J_x^2}+\ex{J_y^2}$ is
smaller than the bound in Eq.~(\ref{4qubit}) thus biseparable pure
states with a $(12)(34)$ partition fulfill Eq.~(\ref{4qubit}).

Now let us take biseparable states with the partition $(1)(234).$ We
will follow similar steps as in the proof of Lemma 2. Let us define
the matrices
\begin{eqnarray}
Q_a&:=& \sigma_a^{(2)}+\sigma_a^{(3)}+\sigma_a^{(4)}; \;\;\;\; a=x,y,\nonumber\\
R&:=&\sum_{l=x,y}
\sigma_l^{(2)}\sigma_l^{(3)}+\sigma_l^{(2)}\sigma_l^{(4)}+\sigma_l^{(3)}\sigma_l^{(4)}.
\end{eqnarray}
Using these matrices we can write
\begin{eqnarray}
\ex{J_x^2}+\ex{J_y^2}&=& 2 +\frac{1}{2}(x_1\exs{Q_x}+y_1\exs{Q_y}+\exs{R})\nonumber\\
&\le&2+\frac{1}{2}\max_{x_1^2+y_1^2\le 1}
\Lambda_{\max}(x_1Q_x+y_1Q_y+R).\nonumber\\ \label{ineqq}
\end{eqnarray}
Now again for finding an upper bound we need the eigenvalues of
$(x_1Q_x+y_1Q_y+R).$ These are
\begin{eqnarray}
\lambda_{1,2}&=& - 2 + X,\nonumber\\
\lambda_{3,4}&=& - 2 - X,\nonumber\\
\lambda_{5,6}&=& 2 + X\pm 2\sqrt{1+X+X^2},\nonumber\\
\lambda_{7,8}&=& 2 - X\pm 2\sqrt{1-X+X^2}, \label{eig}
\end{eqnarray}
where $X=\sqrt{x_1^2+y_1^2}$. Assuming $|X|\le 1$, the upper bound
of the eigenvalues in Eq.~(\ref{eig}) is $3+2\sqrt{3}.$ Thus based
on Eq.~(\ref{ineqq}) we obtain Eq.~(\ref{4qubit}) for biseparable
states with a $(1)(234)$ partition.

Since the measured operators are symmetric under the permutation of
qubits, this also proves that Eq.~(\ref{4qubit}) holds for any
biseparable pure state. Due to the convexity of biseparable states,
it also holds for mixed biseparable states. $\qed$

Criterion Eq.~(\ref{crit3q}) and Theorem 3 have already been used in
the experiment with photons described in
Ref.~\citeonline{Experiment} for detecting multipartite entanglement
in three-qubit and four-qubit systems. Let us now briefly outline
how to detect multipartite entanglement for more than four qubits.
For many qubits detecting multipartite entanglement becomes
difficult with collective observables, since (i) the robustness to
noise is decreasing as the number of qubits are increasing and (ii)
it is very hard to obtain the bound for biseparable states for an
operator expectation value. The first problem can be handled
building entanglement criteria which use higher order moments of the
angular momentum coordinates $J_{x/y}.$ This makes the robustness to
noise somewhat better. The second problem can be overcome, for
example, by using the method applied in
Refs.~\citeonline{TG05,TG05_2}. This makes it possible to find upper
bounds for operator expectation values for biseparable states for
large number of qubits.

Finally, let us discuss how our entanglement conditions
Eqs.~(\ref{crit},\ref{crit3q},\ref{4qubit}) are connected to
superradiance. The left hand side of Eq.~(\ref{crit}) is the same
expression which appears in Eq.~(28) of Dicke's original paper
\cite{Dicke} giving the intensity of the superradiant light during
spontaneous emission in a cloud of atoms. To be more precise, the
light intensity is $I:=I_0\ex{J_x^2+J_y^2+J_z}$ where $I_0$ is the
radiation rate of one atom in its excited state. Criterion
Eq.~(\ref{crit}) shows that if $I/I_0-\ex{J_z}$ is larger than a
bound then the system is entangled. We can also see that there are
separable states [e.g., the state presented in Eq.~(\ref{psixy})]
for which the light intensity scales roughly with the square of the
number of qubits.

\section{Conclusion}

We presented several methods for detecting entanglement in the
vicinity of symmetric Dicke states with multiple excitations. In
particular, we focused on $N$-qubit symmetric Dicke states with
$N/2$ excitations. We showed that they are well suited for
experiments aiming to create and detect multi-partite entanglement.
We presented fidelity-based criteria for detecting genuine
multi-qubit entanglement in the vicinity of these states. We also
considered entanglement criteria based on the measurement of
collective observables. The relation of our entanglement conditions
to superradiance was also discussed.

\section*{ACKNOWLEDGMENTS}

We would like to thank O. G\"uhne, N. Kiesel, C. Schmid, E. Solano,
M.M. Wolf, and H. Weinfurter for useful discussions. We also
acknowledge the support of the European Union (Grants No.
MEIF-CT-2003-500183 and No. MERG-CT-2005-029146), the EU projects
RESQ and QUPRODIS, and the Kompetenznetzwerk
Quanteninformationsverarbeitung der Bayerischen Staatsregierung and
the National Research Fund of Hungary  OTKA under Contract No.
T049234.

\appendix

\section*{APPENDIX: Proof of Eq.~(\ref{ineq})}

Here we present the proof of Eq.~(\ref{ineq}). First let us fix
$N_1$ and look for the maximum of the left hand side of
Eq.~(\ref{ineq}) as the function of $k.$ (Without loss of
generality, we consider $N_1 \le N/2.$) We define
\begin{equation}
g_k:=\bigg(\begin{array}{c}N_1 \\ k\end{array}\bigg)
\bigg(\begin{array}{c}N-N_1 \\ \frac{N}{2}-k\end{array}\bigg).
\end{equation}
Let us look for the $k$ for which it is maximal. For that we compute
the ratio of two consecutive $g_k$
\begin{equation}
\frac{g_{k-1}} {g_{k}}=\frac{k(N/2-N_1+k)}{(N_1-k+1)(N/2-k+1)}.
\label{rat}
\end{equation}
The right hand side of  Eq.~(\ref{rat}) equals $1$ for
$k_m=(N_1+1)/2.$ Thus for $k<k_m$ we know that $g_k/g_{k-1}\ge 1$
while for $k>k_m$  we have $g_k/g_{k-1}\le 1.$ Simple calculation
shows that the integer value for which $g_k$ is maximal is $k=N_1/2$
for even $N_1$ and $k=(N_1 \pm 1)/2$ for odd $N_1.$

Now we know that the maximum of the left hand side of
Eq.~(\ref{ineq}) for a given $N_1$ is
\begin{equation}
h_{N_1}:=\bigg(\begin{array}{c}N_1 \\
\lfloor\frac{N_1}{2}\rfloor\end{array}\bigg)
\bigg(\begin{array}{c}N-N_1 \\
\frac{N}{2}-\lfloor\frac{N_1}{2}\rfloor\end{array}\bigg),
\end{equation}
where $\lfloor x\rfloor$ denotes the integer part of $x.$ We find
that for even $N_1$
\begin{equation}
\frac{h_{N_1}} {h_{N_1\pm 1}} \ge 1.
\end{equation}
Hence $h_{N_1}$ must be maximized for some even $N_1.$ Further
calculation shows that for even $N_1$
\begin{equation}
\frac{h_{N_1}} {h_{N_1-2}}= \frac{N_1-1}{N_1}
\frac{N-N_1+2}{N-N_1+1} \le 1.
\end{equation}
Hence we know that $h_{N_1}$ is maximized by $N_1=2.$ Thus we proved
that the left hand side of Eq.~(\ref{ineq}) is maximized for $N_1=2$
and $k=1.$


\begin{thebibliography}{99}
\bibitem{moregeneral} Note that this condition can also be written
in a more general way with the eigenvalues of the correlation matrix
$C$ defined as $c_{ab}:=\exs{J_a J_b+J_b J_a}/2.$ Assuming
$\lambda_1\ge\lambda_2\ge\lambda_3$, Eq.~(\ref{crit}) can be
reformulated as $\lambda_1+\lambda_2\le N(N+1)/4.$
\end{thebibliography}
\end{document}